\documentclass[twocolumn,showpacs,preprintnumbers,amsmath,amssymb]{revtex4}

\usepackage{graphicx}
\usepackage{dcolumn}
\usepackage{bm}

\begin{document}

\preprint{---}

\title{Do we really need the weight force?}

\author{Jos\'{e} M. L. Figueiredo}
 \email{jlongras@ualg.pt}
\affiliation{Departamento de F\'{\i}sica da Faculdade de Ci\^{e}ncias e Tecnologia da Universidade
do Algarve, Campus de Gambelas, 8005-139 FARO, Portugal
}%

\date{\today}

\begin{abstract}
In 1901 the third Conf\'{e}rence G\'{e}n\'{e}rale des Poids et Mesures (CGPM) defined the weight of
a body as ``the product of its mass and the acceleration due to gravity". In practical terms the
weight force corresponds to the gravitational force. However, this gravitational definition of the
weight lacks logic from the perspective of present knowledge and can be misleading. In a space
traveling time, weight associated concepts such as ``true weight", ``apparent weight",
``weightlessness", ``zero gravity", ``microgravity", are ambiguous and strong deceptive. The
superfluousness or ambiguousness of these concept would be removed if an operational definition of
the weight of a body as \textbf{the force the body exerts on its support or suspender} is adopted.
In this definition the weight force does not act on the body and it is not necessary to describe
the body's sate of motion. This avoids the imprecision of the concepts such as ``true weight" and
``apparent weight" of a body, for example. This paper discusses the need of the weight force of a
body concept in Physics, and asks for a re-exame of the CGPM definition to improve the physics
teaching and learning.


\end{abstract}



\maketitle

\section{Introduction}

Everyone who teaches introductory physics is used to deal with the confusion students experience
when first confronted with the differences between concepts of mass and weight, and with the
ambiguities of the notions related to the weight definition based on the gravitational force such
as ``true weight", ``apparent weight", ``weightlessness", ``zero-gravity", and ``vertical". Several
studies have shown college and first year university students still have some misconception about
the basic physics related to the meaning of weight and have difficulties applying it in
imponderability or accelerated environments \cite{Galili}\cite{GurelAcar}\cite{Sharma}. This is
evident when they are asked to explain what happens if some everyday events such as walking were to
take place in imponderability. The students often think ``microgravity", ``zero gravity", or
``weightlessness" situations refer to events occurring outside the Earth's or other celestial
body's gravitational influence. They are surprised to find out that during a typical shuttle flight
mission in an orbit at an altitude of 400 km the gravitational force of the Earth is only 12\% less
than at the Earth's surface.

In the literature there are several definitions of weight based on the gravitational force: ``the
weight is the Earth gravitational force" \cite{Serway}; ``the force exerted by the Earth on an
object is called the weight of the object" \cite{Serway}; ``the weight of a body is the total
gravitational force exerted on the body by all other bodies in the universe" \cite{YoungFreedman}.
Frequently, the weight is considered a fundamental property of matter under the influence of a
gravitational field. Furthermore, in 1901 the Conf\'{e}rence G\'{e}n\'{e}rale des Poids et Mesures
declared ``the weight of a body is the product of its mass and the acceleration due to gravity"
\cite{3CGPM}.

These gravitational based definitions of weight are widely used in despite of the fact they are not
entirely satisfactory at the present knowledge and the number of ambiguities associated to the
gravitational definition and on the meaning of weight and weight-related concepts are even
misleading, leading to several misconceptions which can contribute to widening the gap between what
is taught and what is learned by the students.

Very few authors adopt the alternative operational definition of weight of a body as ``the force
which a body exerts on its support or suspender that prevents its free fall" \cite{MarionHornyak}.
This definition is in agreement with our daily experience and with the present knowledge. Following
this definition the weight of a body is a force that results always from the direct contact of the
body with other body, i.e., the weight force is a contact force.

This paper discusses weight, microgravity, weightlessness, vertical, up and down concepts, and the
necessity of reviewing the weight concept/definition together with the advantages of the adoption
of the operational definition of weight or its abandonment.

\section{Acceleration due to the gravity}

In the frame of the classical physics, the force of gravity is a long-range force, and, as far we
know, cannot be shielded \cite{gravitationalshielding}. In practical situations it is independent
of the state of motion of the objects. The acceleration due to the gravity corresponds to the
acceleration of the motion of a body as a result of the gravitational force, and in a given instant
equals the ratio of the gravitational force and the body amount of matter.

Accordingly the General Theory of Relativity the gravity corresponds to a modification (curvature)
in the space-time continuum caused by a concentration of mass or energy, that is, the space-time
geodesics surrounding substantial mass or energy are curved lines and the bodies go through some
form of curved orbital path.

\subsection{Gravity under newtonian physics}\label{gravity-newton}

Since Isaac Newton presented the law of Universal Gravitation it is well accepted that the
gravitational interaction is universal and depends only on the body's quantity of matter and the
distance between their centers of mass. Following the works of Kepler and Galileu, Newton concluded
that the Earth's force of gravity \(\vec{F}_{g}\) exerted on our bodies or other mass \(m\) owing
to their gravitational interaction is given by
\begin{equation}\label{FgTm}
\vec{F}_{g}=-\frac{GMm}{|\vec{r}|^{3}}\vec{r}=\vec{\Gamma}m,
\end{equation}
where \(G\) is the gravitational constant (6.67\(\times10^{-11}\) N m\(^{2}\)kg\(^{-2}\)), \(M\) is
the Earth's mass, \(\vec{r}\) is the position vector of the body center of mass relatively to the
Earth center of mass. The gravitational force the Earth exerts on the body \(\vec{F}_{g}\) can be
written as the product of the body's mass and the local acceleration due to the gravity,
\(\vec{F}_{g}=m\vec{g}\); the vector \(\vec{g}=-\frac{GM}{|\vec{r}|^{3}}\vec{r}\) corresponds to
the body's acceleration due to the Earth gravitational field \(\vec{\Gamma}\), and its magnitude is
approximately equal to 9,8 m s\(^{-2}\) at sea level. The intensity of the force of gravity can be
measured with the aid of a dynamometer (or a spring scale), provided that the body and the
dynamometer are at rest relatively to the Earth.

\begin{figure}[hbt]
\begin{center}
\includegraphics{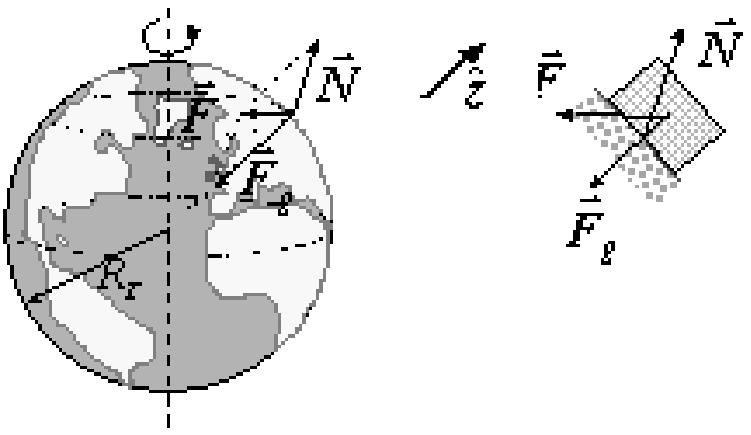}
\end{center}
\caption{\label{Fig1} Forces acting on a body at the Earth's surface for a given latitude:
\(\vec{F}\) is the net force due to the force of gravity \(\vec{F}_{g}\) acting on the body and to
the reaction of the Earth's surface \(\vec{N}\) acting on the body. \(\vec{N}\) and \(\vec{F}_{g}\)
do not constitute an action-reaction pair because they are applied to the some body.}
\end{figure}

Let us consider a body at rest on the surface of the Earth at a given latitude, Fig. \ref{Fig1}.
The body is acted upon by two forces: the force of gravity \(\vec{F}_{g}\) pointing towards the
center of the Earth and the force of the reaction of the Earth's surface (the support reaction
force) \(\vec{N}\), whose direction is determined not only by the force of gravity, but also by the
spinning of the Earth around its axis. Accordingly the second law of dynamics
(\(\vec{F}=d(m\vec{v})/dt\), where \(\vec{v}\) is the velocity of the mass \(m\)), the resultant
force \(\vec{F}\) of these two forces ensures the daily rotation of the body along the local
parallel. As a consequence the directions of \(\vec{N}\) and \(\vec{F}_{g}\) (given by equation
\ref{FgTm}) do not coincide. The direction of the measured \(\vec{F}_{g}\) and \(\vec{g}\) (so the
direction of \(\vec{N}\)) differs from the direction towards the center of the Earth given by
equation \ref{FgTm}, except at the poles and equator, by an angle whose the maximum amplitude is
less than 0.1\(^{0}\). In addition with the exception at the poles, a scale or a dynamometer
measures less than the gravitational force given by equation \ref{FgTm} because the net force
needed to provide the centripetal acceleration necessary to ensures the body keeps up with the
daily rotation of the Earth: assuming a spherically symmetric and homogenous Earth the sensed
acceleration of gravity is about 0.03 m s\(^{-2}\) (0.35\% of \(g\) given by
\(\vec{g}=-\frac{GM}{|\vec{r}|^{3}}\vec{r}\)) less at the equator than at the poles. Furthermore,
the variation of density and the surface irregularities of the Earth give rise to local changes in
the gravitacional field and to the vector \(\vec{g}\).

Nevertheless, throughout the rest of the text we will consider the Earth as an homogenous sphere
and the effects of its rotation around its axis and the translation around the Sun or other motions
will be neglected because the values of the linear and the angular accelerations acquire by a body
due to these effect are very small when compared with the acceleration due to the gravity. For
simplicity, the Earth will be considered a frame of reference at rest during the characteristic
time of the phenomena analyzed here. The effect of the atmosphere and the gravitational influences
of other celestial bodies are also neglected.

\subsection{Gravity under the theory of general relativity}

The General Theory of Relativity addresses the problems of gravity and that of nonuniform or
accelerated motion. In one of his famous conceptual experiments Einstein concluded that it is not
possible to distinguish between a frame of reference at rest in a gravitational field and an
accelerated frame of reference in the absence of a significant gravitational field:
Einstein's principle of equivalence. From this principle of equivalence Einstein moved to a
geometric interpretation of gravitation: the presence of very large mass or a high concentration of
energy causes a local curvature in the space-time continuum. The space-time geodesics become curved
lines, that is, the space-time curvature is such that the body paths are no longer straight lines
but some form of curved orbital paths. Maintaining the classical view of the gravitation we can
associate to the body's curved path motion a \emph{centripetal acceleration} that is referred as
the acceleration due to gravity.

\section{Operational definition of the force weight of a body}

What humans and matter experience as weight is not the force of gravity. What they experience as
weight is actually the consequence of the reaction of the ground (or whatever surface they are in
contact with or hang up) pushing ``upwards" against them to counteract the force they are exerting
on the surface - \emph{the weight force of the body}. A good evidence of this is given by the fact
that a person standing on a scale moving up and down on his toes does see the indicator moving,
telling that the measured force is changing while the gravity force, that depends only on the
person's and the Earth' masses and the distance between their centers of mass, does not vary to
induce such clear observable changes on the scale meter. Another evidence happens when going
towards the Earth surface in an elevator: one experiences a greater strain in the legs and feet
when the elevator is stopping than when it is stationary or moving with constant velocity because
the floor is pushing up harder on the feet.

\subsection{Weight of a body at rest}

\begin{figure}[hbt]
\begin{center}
\includegraphics{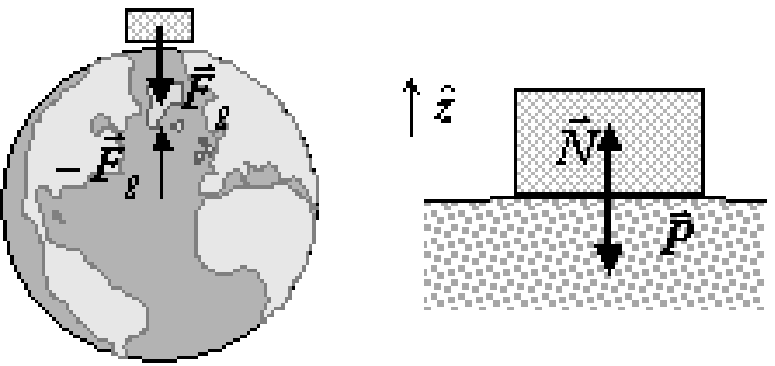}
\end{center}
\caption{\label{Fig2}a) \(\vec{F}_{g}\) and \(-\vec{F}_{g}\) is the action-reaction force pair due
to the gravitational interaction between the body and the Earth. b) \(\vec{P}\) and \(\vec{N}\) is
the action-reaction force pair due to the interaction between the body and the Earth surfaces:
\(\vec{P}\) is action of the body on the support (the weight force of the body) and \(\vec{N}\) is
the reaction of the support's surface to the body's action \(\vec{P}\) on the support surface.}
\end{figure}

Consider a body at rest on the surface of the Earth, Fig. \ref{Fig2}a \cite{MarionHornyak}. In this
situation the body experiences a force \(\vec{F}_{g}\) due to gravitational pull of the Earth. The
reaction force to this force is \(-\vec{F}_{g}\) and corresponds to the gravitational force exerted
on the Earth by the body. The force pair \(\vec{F}_{g}\) and \(-\vec{F}_{g}\) constitutes an
action-reaction pair, and consists of the force \(\vec{F}_{g}\) that acts on the body and the force
\(-\vec{F}_{g}\) that acts on the Earth.

The tendency of the body to accelerate towards the center of the Earth due to \(\vec{F}_{g}\) must
give rise to a force \(\vec{P}\) acting on the Earth surface: \(\vec{P}\) is the force exerted by
the body on the Earth surface, Fig. \ref{Fig2}b. If the body exerts on the Earth surface a force
\(\vec{P}\), the Earth solid surface reacts exerting a force \(\vec{N}\) on the body that balances
the force \(\vec{P}\), Fig. \ref{Fig2}b. The force \(\vec{N}\) is called the normal force and is
the reaction to \(\vec{P}\) and we have \(|\vec{N}|=|\vec{P}|\). The action \(\vec{P}\) the body
exerts on the Earth (or other body) surface corresponds to the force \textbf{weight of the body}
(that is the operational definition of weight).

Hence the body experiences no acceleration (it is at rest), the net force due to the two forces
acting on the body, \(\vec{F}_{g}\) towards the center of the Earth and \(\vec{N}\) outwards, is
null (Newton's second law of dynamics). Therefore, \(\vec{F}_{g}\) and \(\vec{N}\) are equal in
magnitude, opposite in orientation and have different application points. Although in this case
\(|\vec{N}|=|\vec{F}_{g}|\) the normal force \(\vec{N}\) is not the reaction to the gravitational
force \(\vec{F}_{g}\) because this two forces act on the body (as said previously \(\vec{N}\) is
the reaction to \(\vec{P}\)). Similarly, because the Earth experiences no acceleration there are
two equal and directly opposite forces acting on the Earth, \(-\vec{F}_{g}\) applied on the Earth's
center of mass and \(\vec{P}\) applied on the Earth surface in contact with the body, with
\(|\vec{P}|=|-\vec{F}_{g}|\).

Consider now the body is placed on (or hung on) a dynamometer-scale. When the body is placed on the
scale platform the dynamometer spring is compressed or extended (depending on the scale) and its
deformation is communicated to a calibrated dial read out. The body exerts an action \(\vec{P}\) on
the scale platform and through it on the spring. The scale dial reads the magnitude of the force
\(\vec{P}\) exerted \textbf{by} the body surface \textbf{on} the scale platform. By Newton's third
law of dynamics the scale platform reacts exerting a opposing force \(\vec{N}\) \textbf{on} the
body surface: both forces have the same magnitude, \(|\vec{P}|=|\vec{N}|\), and direction but
opposite orientations. It is the force  \(\vec{N}\) that prevents the  body free fall towards the
center of the Earth. The weight force is applied not to the body being considered itself, but to
the scale platform.

If the scale \textbf{is at rest} relatively to the Earth then, as previously, the magnitude of the
reaction force the support exerts on the body 
equals the magnitude of the force of gravity acting on the body. Hence \(|\vec{N}|=|\vec{F}_{g}|\)
it results that \(|\vec{P}|=|\vec{F}_{g}|\). The weight force and the force of gravity magnitudes
are also equal \textbf{in the case of uniform and rectilinear motion} of the scale and the body in
a reference frame associated with the Earth.

What happens when the scale and body are accelerating in relation to a frame of reference on the
Earth? This is the case, for example, of an elevator during stopping or starting. What is the scale
reading in these situations? During a sudden change in the support motion (on starting or braking,
for example) does remain valid the equality \(|\vec{P}|=|\vec{F}_{g}|\)?

\subsection{Weight of a body in an accelerated frame of reference}

Consider that a body of mass \(m\) is standing on a bathroom-type scale fixed in the floor of an
elevator. When the elevator moves with an acceleration \(\vec{a}\), in accordance with Newton's
second law of dynamics and as long as the body and the scale surfaces are in contact, the body
moves together with the elevator and scale with the acceleration \(\vec{a}\) under the action of
two forces: the force of gravity \(\vec{F}_{g}\) and the scale surface reaction force \(\vec{N}\)
due to the body surface action \(\vec{P}\) on the scale. From Newton's second law of dynamics,
\(\vec{F}=d(m\vec{v})/dt\), and assuming the body's mass does not vary, the resultant of the forces
acting on the body must equal the product of its mass by its acceleration, which is the elevator
acceleration \(\vec{a}=d\vec{v}/dt\), that is,
\begin{equation}\label{000}
    m\vec{a}=\vec{F}_{g}+\vec{N}.
\end{equation}
Since the body's weight \(\vec{P}\) and the reaction force of the scale platform \(\vec{N}\)
constitutes an action-reaction pair, `\(\vec{P}=-\vec{N}\)´, equation \ref{000} can be rewritten as
\begin{equation}\label{00}
    \vec{P}=m\vec{g}-m\vec{a}.
\end{equation}
It is important to bear in mind that in the operational definition the \emph{weight of the body}
\(\vec{P}\) corresponds to the body's action force on the support, that is, the weight force of a
body does not act on the body, Fig. \ref{Fig2}b. \textbf{As a consequence to describe the motion of
a body we do not have to consider the weight force of the body.}

Equation \ref{00} allows us to determine the action a body exerts (the weight force of the body) on
its support or suspender. Depending on the orientation of the acceleration \(\vec{a}\), several
situations may occur \cite{MarionHornyak}\cite{FigueiredoJ-AJP}: i) \(|\vec{P}|<mg\); ii)
\(|\vec{P}|=0\); iii) \(|\vec{P}|>mg\). A classical case of study is the movement of an elevator
with a passenger standing on a dynamometer scale (see for example
\cite{MarionHornyak})\cite{FigueiredoJ-AJP}.

\subsection{Notions of vertical, upward and downward}

Currently, the vertical is defined as the direction of the plumb line that at the Earth surface and
at rest or on uniforme and rectilineal motion coincides with the direction of the gravity force
adopting the consideration of section \ref{gravity-newton}. However, a human being or other living
creature feels equilibrated in the direction of its weight force, and the vertical direction and
downward orientation correspond to the direction and to the orientation of the weight force,
respectively. From equation \ref{00} one can conclude the vertical and the up and down orientations
depend essentially on the body state of motion characteristics. Contrary to what is many times
stated the notions of vertical and up/down are not determined uniquely by the gravitational force.
The vertical is always the direction of the weight of the body and ``down" corresponds to the
weight force orientation. In accordance equation \ref{00}, to stay in equilibrium during a bus
starting movement we stoop forward and when it starts stopping we lean backwards. In these
situations our vertical is oblique and to not lose one's balance we align with the new vertical
defined by the new direction of \(\vec{P}\).

\section{The weight force in a weak gravitational field}\label{p-verus-g}

The body's action force, the weight force, ``appears" whenever the body's surface is constrained to
interact directly with the surface of another body. The weight of the body is opposite to the
normal force (reaction force) exerted by the surface where the body stands on or is in contact
with, which prevents it from moving through or away of the other body surface. The body action
force or its absence (weightlessness) does not depende of the existence of a gravitational field in
the region of the space where it is staying. Considere a spaceship in a region of the Universe
where the gravitational field is very small. The bodies in the interior of a spaceship traveling in
this region with uniforme and rectilinear motion would experience zero weight, because they are in
a situation of effective zero-gravity  and with uniforme and rectilinear motion. Any spring-scale
in contact and moving with them measures no weight because the objects are not constrained to
contact their surfaces to originate the normal forces (zero action - zero weight - gives rise to no
normal force).

Lets now considere the spaceship turns on its engines. In the case of a spacecraft accelerating by
firing its rockets the thrust force is applied to the back end of the rocket by the gas escaping
out of the back. The rockets thrust force is transferred to each object in the spaceship through
either pressure or tension giving rise to the bodies action (weight force) on their supports or
suspenders and the bodies in the interior of the veicule do not experience weightlessness. We can
conclude that the weight force in fact does not depend on the presence of a gravitational field.
Indeed, according to Einstein's Principle of Equivalence the bodies in a space veicule with an
acceleration \(\vec{a}\) in the absence of a gravitational field behave as the spaceship was at
rest or with constant velocity in a gravitational field with acceleration due to gravity
\(\vec{g}=-\vec{a}\).

From equation \ref{00} results that if the spaceship is accelerating uniformly out of the influence
of a significant gravitational field, that is, \(|\vec{g}|<<|\vec{a}|\), the weight of a body in
the vehicle is
\begin{equation}\label{0000}
    \vec{P}=-m\vec{a},
\end{equation}
that is, the weight force is opposite to the net force acting on the body and it is equal to the
product of the body's acceleration and mass, \(m\vec{a}\). Taking in account the considerations
made and the equation \ref{0000} the weight of a body or the weightlessness state has nothing to do
whether the body is under the influence of a gravitational field or not. In conclusion, the forces
the bodies exert on their supports (weight forces) or their absence do not require the presence of
a gravitational field. In the case of the presence of a gravitational field the force the bodies
exert on their supports depends also on characteristics of their relative motion.

\section{Imponderability and microgravity}\label{imp-micro}

In free fall, that is, when \(\vec{g}=\vec{a}\) all parts of an object accelerate uniformly and
thus a human or other body would experience no weight, assuming that there are no \emph{tidal
forces} \footnote{The tidal force are secondary effects of the forces of gravity due to Earth
inhomogeneities and to the other celestial objects gravity.}. The experience of no weight, by
people and objects, is known as imponderability, weightlessness or \emph{zero gravity}, although
\emph{micro-gravity} is often used to describe such a condition. Excluding spaceflight (orbital
flight), weightlessness can be experienced safely only briefly, around 30 seconds, in an airplane
following a ballistic parabolic path. In spaceships the state of imponderabilidade or
weightlessness can be experienced for extended periods of time if the ship is outside the Earth's
or other planet's atmosphere and as long as no propulsion is applied and the veicule is not
rotating about its axis because the bodies in its interior are not constrained to be in contact
with other bodies or with the station walls or floors. In particularly, the astronauts are not
pulled against the station pavement and therefore their bodies actions on the surface of the
station are null. In real free fall situations the tidal effects of the gravity on the bodies,
although small, are equivalente to a small acceleration  and the bodies are said to be in a
``microgravity" environment because the weightlessness sensation is not complete.

The sate of imponderability experienced in orbiting spacecrafts is not as consequence of the small
value of the acceleration due to the gravity because the distance from the Earth.
Weightlessness is a consequence of the body and the spaceship accelerations to be equal and only
due to gravity. The gravity acts directly on a person and other masses just like on the vehicle and
the person and the floor are not constrained toward each other. On the contrary, contact forces
like atmospheric drag and rocket thrust first act on the vehicle, and through the vehicle on the
person. As a consequence of this contact forces the person and the floor are pushed toward each
other giving rise to the weight force.

As mentioned the term microgravity is usually used instead of weightlessness to refer the
environment within orbiting spacecraft. The use of the term microgravity without specifying its
exact meaning can strengthen the misconceptions associated to weight and gravitational force
because the term ``micro" could lead to the idea that acceleration due to gravity is very small
because the distance from Earth. To the contrary, the acceleration of the gravity due to the Earth
gravitational interaction is around 8.4 m s\(^{-2}\) at 400 km of altitude (88.8\% of it value at
the Earth's surface). Even its value at the distance of the Moon orbit is 2.63\(\times10^{-3}\) m
s\(^{-2}\), although in these regions the acceleration due to Sun's gravity is near twice this
value (\(\approx5.8\times10^{-3}\) m s\(^{-2}\)). True Earth's microgravity,
\(g\simeq1\times10^{-5}\) m s\(^{-2}\), can be only experienced at locations away from the Earth as
far as almost 17 times the Earth-Moon distance.

The term microgravity is used by the scientists to characterize the residual acceleration
experienced by the bodies in the interior of the spacecraft as a consequence of forces between the
bodies within the spaceship, the body and the spacecraft, the gravitational tidal forces and the
atmosphere dragging force. These forces induce in the bodies acceleration of intensities of some
\(\mu\)m s\(^{-2}\), giving rise to the use of the term ``microgravity". For uncrewed spaceships
free falling near the Earth it is possible to obtain 1 \(\mu g\); for crewed missions is difficult
to achieve less than 100 \(\mu g\) \footnote{``\emph{Falling upwards: how to create microgravity},"
http://w\newline ww.esa.int/esaHS/ESATRRVRXLC\underline{ }research\underline{ }0.html/; ``\emph{The
Mathematics of Microgravity}," http://exploration\newline .grc.nasa.gov/combustion/pdf\underline{
}files/math\underline{ }of\underline{ }micro.pdf.}. The main reasons are: i) the morphology of the
Earth induces local gravitational variations; ii) the gravitational effects of the other celestial
bodies, especially the Moon and the Sun, which depend on their relative position to the Earth; iii)
the acceleration due to gravity decreases one part per million for every 3 m increase in altitude
(in an orbiting spaceship the required centripetal force and hence the acceleration due to gravity
is higher at the far side than at the nearest side of the ship relatively to the Earth); iv)
although very thin, at for example 400 km of altitude, the atmosphere gradually slows the
spacecraft.

\section{Living in microgravity environment }\label{imp-micro}

The term microgravity is more appropriate than ``zero weight" or ``zero-gravity" in the case of
orbiting spacecrafts because weightlessness is not perfect. The term microgravity does not mean the
acceleration due to gravity was strongly reduced but solely that its effects on the bodies within
the vehicle were substantially reduced.

As already mentioned the bodies in the interior of a spacecraft orbiting a celestial body, such as
the International Space station (ISS) around the Earth, are in a state of imponderability
experiencing weightlessness as they do not exert any contact action on the other bodies because all
the bodies are subjected only to the gravity force being pulled towards the Earth with the same
acceleration, the acceleration due to gravity. The weightlessness present several challenges to the
human organism which is prepared to live in a gravity environment and also makes several of the
mundane human actions, such as walk, virtual impossible.

Because in the interior of the station there are no upward and downward convection currents of
particles and gases that leads to several effects on the human breathing system. The weightlessness
also interferes with cardiovascular system with the heart beating faster because there is less
resistance to the blood flow. This can also lead to the muscles atrophy, blood pump system
malfunction and difficult breathing.

It is not possible to walk in weightlessness environment because the astronauts feet are not
constrained to the station pavement and their feet action (weight force) on the surface of the
station is null. Hence there is no normal reaction force therefore the friction force is zero. At
the Earth surface it is the friction force between the pavement and the astronauts feet that gives
rise to the reaction force needed to walk.

Several plans have been proposed to create ``artificial gravity" in orbiting devices. The most
popular plan to produce ``artificial gravity" in vehicles designed to remain in orbit or stay in
out space for a long period of time are to set the spaceship into rotation with an angular velocity
\(\omega\) around its central axis. The bodies at any point distant by \(r\) from the rotation axis
will experience a centripetal acceleration \(a=\omega^{2}r\). The weight of the bodies on the outer
rim of the spaceship with radius \(r\) opposes the centripetal force.  The weight force intensity
is then given by \(P=m\omega^{2}r\).

Although this process could be used to simulate gravity it wouldn't be exactly the same. One
problem is that across the radius of a spaceship, \(g\) levels change rapidly, and different parts
of a human body will feel considerable distinct acceleration levels. That is not quite what we
experience in Earth's gravitational field.

\section{Conclusion}

The identification of weight force as the force of gravity is misleading and lacks logic from the
perspective of the present knowledge. In the operational definition discussed the weight of a body
is the action force the body exerts on the surface of another body that it is in contact with, and
depends on their relative motion. The current meaning of weight is in fact the gravitational force
and to describe the motion of a body we do not need to consider the weight force if operational
definition is adopted. In fact, having in mind  the concept of weight is not fundamental the
Physics teaching and learning would benefit if the use of the vocable weight is avoided. One
advantage would be to get rid of the common sense identification between mass and weight force
concepts in the class room.
It is expected that this analysis will motivate physics scientific community, as well as
instructors and authors, to contribute to the review of the concept of weight either adopting the
operational definition or considering to lay it aside.

\section*{Acknowledgments}

The author is grateful to Professor Robertus Potting, Dr. Paulo S\'{a}, Dr. Jos\'{e} Rodrigues, and
Dr. Alexandre Laugier for their comments and manuscript revision. The author acknowledges the
improvements that resulted from further discussions with other colleagues.

\end{document}


\preprint{GF RBEF}

\title[The concept of the weight: the operational definition as a contact force]
{Precisámos mesmo da força peso?}




\author{Jos\'{e} M. L. Figueiredo\footnote{E-mail: jlongras@ualg.pt}}
\address{Departamento de F\'{\i}sica da Faculdade de Ci\^{e}ncias e Tecnologia da Universidade do Algarve, Campus de Gambelas, 8005-139 FARO, Portugal}

\begin{abstract}

Com raras excepções, nos manuais de Física para os ensinos secundário e universitário e em muitas
publicações científicas, a força ``peso de um corpo" é definida como a força da gravidade ou uma
consequência exclusiva desta. 
Estas definições carecem de lógica do ponto de vista do conhecimento actual e podem ser mesmo
enganadoras para o estudante.  A definição operacional da força ``peso do corpo" como a força que o
corpo exerce no seu suporte tem a vantagem de eliminar a necessidade de vários conceitos associados
à definição gravitacional ou, pelo menos, a ambiguidade de alguns deles. Neste texto discute-se o
conceito da força ``peso" e apresentam-se exemplos das vantagens da sua redefinição e, em
particular, do abandono da expressão ``peso do corpo" uma vez que a noção de força ``peso do corpo"
não é fundamental em Física.


\end{abstract}
\date{\today}


\maketitle

\section{Introdução}

Quem lecciona disciplinas introdutórias de Física está acostumado a lidar com a confusão que os
estudantes fazem quando confrontados com os conceitos de massa e de força ``peso", e com as
ambiguidades das noções associadas à definição da força ``peso do corpo" baseada na força da
gravidade. Vários estudos indicam que os estudantes do secundário e da universidade mostram
dificuldades em aplicar os conceitos relacionados com a força ``peso" em situações de
imponderabilidade  e em sistemas com movimento acelerado
\cite{Galili}\cite{GurelAcar}\cite{Sharma}.
Com frequência, os alunos consideram que as situações correntemente designadas como
``gravidade-zero", ``microgravidade" ou ``sem peso" se referem a eventos que ocorrem fora da
influência gravitacional da Terra ou de outro corpo celeste e ficam, muitas vezes, surpreendidos
quando percebem que durante uma missão do vaivém espacial, por exemplo, a aceleração devida à
gravidade a que ficam sujeitos os astronautas é apenas 12\% inferior à que experimentam à
superfície da Terra.


Na literatura encontram-se várias definições da força ``peso" de um corpo baseadas na força da
gravidade: ``o peso" é a força gravitacional da Terra", ``a força exercida pela Terra num objecto é
chamada ``peso" do objecto" \cite{Serway}; ``o peso" de um corpo é a força gravitacional total
exercida no corpo por todos os outros corpos do Universo" \cite{YoungFreedman}. Estas definições
são compatíveis com a decisão da terceira reunião da Conferência Geral de Pesos e Medidas (CGPM),
em 1901, que definiu ``peso de um corpo" como o produto da sua massa pela aceleração devida à
gravidade \cite{3CGPM}.

As definições da força ``peso" baseadas na gravidade e os conceitos com ela relacionados são
largamente usadas nas aulas e nos livros de texto de Física, especialmente nas escolas secundárias
e nas disciplinas introdutórias de Física do ensino superior, apesar de serem insatisfatórias e
ambíguas, tendo em conta o conhecimento actual, e são, muitas vezes, origem de confusão.
%
Note-se, por exemplo, a imprecisão e a falta de clareza das noções de ``peso verdadeiro", de ``peso
aparente", de ``sem peso", de ``gravidade zero" e de ``microgravidade".
A ambiguidade dá origem a equívocos que podem facilmente contribuir para o alargamento do fosso
entre o que se pretende ensinar e o que é aprendido pelos estudantes.

Poucos autores adoptam a definição operacional da força ``peso de um corpo": ``força que o corpo
exerce no suporte a que está ligado" \cite{MarionHornyak}. Nesta definição operacional, a força
``peso do corpo" é uma força de ligação tal como, por exemplo, a força de atrito ou a tensão numa
corda. Este artigo discute o conceito de força ``peso do corpo", das noções de gravidade-zero, de
microgravidade, de vertical e de ``para baixo/cima", e as vantagens pedagógicas da adopção da
definição operacional da força ``peso" e/ou o abandono do termo força ``peso" em Física.


\section{Aceleração devida à gravidade}

No modelo da Física Clássica a força da gravidade é uma força de longo alcance entre os corpos com
massa e, tanto quanto se sabe, não pode ser blindada \cite{gravitationalshielding}. Em situações
convencionais é independente do estado de movimento dos corpos. A aceleração devida à gravidade é a
aceleração do movimento de um dado corpo material resultante apenas da acção da força da gravidade
sobre esse corpo e, num dado instante, corresponde à razão entre a força da gravidade e quantidade
de matéria do corpo.

Na descrição dada pela Teoria Geral da Relatividade um campo gravitacional corresponde a
modificações da geometria do espaço-tempo provocada por uma concentração de massa ou de energia. As
geodésicas do espaço-tempo são linhas curvas e os corpos descrevem órbitas correspondentes a estas
linhas curvas.


\subsection{Gravidade segundo a Física Newtoniana}

Desde que Isaac Newton propôs a lei da Gravitação Universal que se aceita que a interacção
gravitacional entre massas é universal e apenas depende das quantidades de matéria em interacção e
da distância que separa os seus centros de massa. A partir dos trabalhos de Kepler e de Galileu,
Newton concluiu que a força da gravidade \(\vec{F}_{g}\) que a Terra exerce num corpo de massa
\(m\), consequência da interacção gravitacional do corpo com o planeta Terra, é dada por
\begin{equation}\label{FgmM}
\vec{F}_{g}=-\frac{GMm}{|\vec{r}|^{3}}\vec{r}=\vec{G} m,
\end{equation}
onde \(G\) representa a constante gravitacional (6,67\(\times10^{-11}\) Nm\(^{2}\)kg\(^{-2}\)),
\(M\) é a massa da Terra, \(\vec{r}\) é o vector posição com origem no centro de massa da Terra e
extremidade no centro de massa do corpo.
A força da gravidade que a Terra exerce num corpo \(\vec{F}_{g}\) pode ser escrita como o produto
da massa \(m\) do corpo pela aceleração
devida à gravidade \(\vec{g}\), \(\vec{F}_{g}=m \vec{g}\); 
a grandeza \(\vec{g}=-\frac{GM}{|\vec{r}|^{3}}\vec{r}\) corresponde à aceleração que o corpo
adquire quando sujeito unicamente à acção do campo gravitacional \(\vec{G}\). Junto à superfície da
Terra \(\vec{g}\) é praticamente constante; a intensidade de \(\vec{g}\) ao nível do mar é,
aproximadamente, 9,8 m s\(^{-2}\).

\begin{figure}[hbt]
\begin{center}
\epsfbox{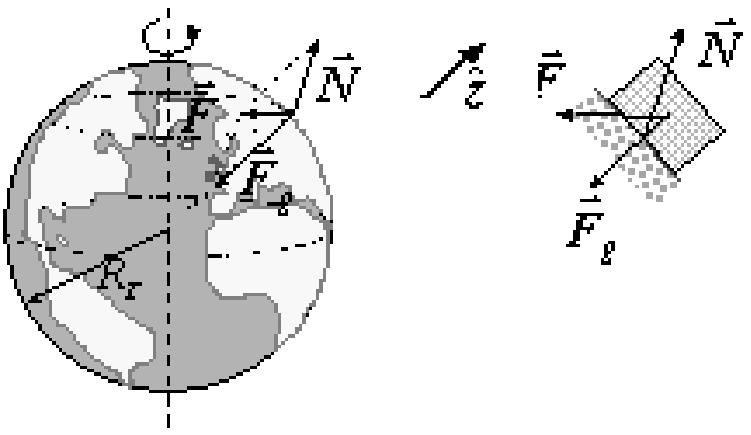}
\end{center}
\caption{\label{Fig1} Forças que actuam no corpo em repouso na superfície sólida da Terra, a uma
dada latitude: a força \(\vec{F}\) é a resultante da força da gravidade \(\vec{F}_{g}\) que actua
no corpo e da acção da superfície da Terra \(\vec{N}\) no corpo.}
\end{figure}

A intensidade da força da gravidade pode ser medida com auxílio de um dinamómetro ou de uma
balança-dinamómetro, assegurando que o corpo e o dinamómetro estão em repouso em relação à Terra.
Esta afirmação requer alguns esclarecimentos.
Considere-se um corpo em repouso relativamente à superfície da Terra a uma dada latitude, Fig.
\ref{Fig1}. O corpo é actuado por duas forças: a força da gravidade \(\vec{F}_{g}\), apontando para
o centro da Terra, e a força de reacção \(\vec{N}\) que a superfície da Terra (ou do suporte)
exerce sobre a superfície do corpo. A direcção de \(\vec{N}\) é determinada pela força da gravidade
e pela rotação da Terra em torno do seu eixo. Tendo presente a segunda lei de Newton da Dinâmica (a
resultante das forças que actuam num corpo é igual à taxa temporal de variação do momento linear do
corpo, \(\vec{F}=d(m\vec{v})/dt\)), a resultante \(\vec{F}\) destas duas forças assegura a rotação
diária do corpo segundo o paralelo que passa pela posição deste. Como consequência apenas deste
efeito, as direcções de \(\vec{F}_{g}\) e de \(\vec{g}\) ``medidas" diferem ligeiramente da
direcção do centro da Terra - excepto nos pólos e no equador - em um ângulo cuja amplitude é
inferior 0,1\(^{0}\). Verifica-se também que, devido à aceleração centrípeta do corpo, a
intensidade de \(\vec{F}_{g}\) indicada na balança ou no dinamómetro é inferior, excepto nos pólos,
ao valor dado pela equação \ref{FgmM}. Acresce ainda que as \emph{irregularidades} da superfície e
as variações de densidade nas diferentes regiões que constituem a Terra dão origem a um campo
gravitacional não exactamente central pelo menos nas proximidades da superfície da Terra e,
portanto, a variações na direcção e na intensidade de \(\vec{g}\).

Ao longo do resto do texto considera-se que a Terra é uma esfera homogénea e desprezam-se os
efeitos da rotação em torno do seu eixo e da translação em torno do Sol, e de quaisquer outros
movimentos, devido aos pequenos valores das acelerações linear e angular da Terra quando comparados
com aceleração devida à gravidade ou a outras forças aplicadas. Isto é, a Terra é considerada em
repouso durante os tempos característicos dos fenómenos aqui analisados. É desprezado também o
efeito da atmosfera no corpos.

\subsection{Gravidade segundo a Teoria da Relatividade Geral}

A Teoria da Relatividade Geral trata a gravidade e o movimento acelerado. Numa das suas famosas
experiências conceptuais, Einstein concluiu que não é possível distinguir entre um referencial em
repouso num campo gravitacional homogéneo e um referencial acelerado na ausência de um campo
gravitacional: princípio de equivalência de Einstein.
A partir deste princípio de equivalência, Einstein propõe uma interpretação geométrica da
gravitação: a presença de massa ou de energia provoca a curvatura do contínuo espaço-tempo, e as
geodésicas deixam de ser linhas rectas e passam a ser percursos curvos.
Numa descrição clássica deste efeito podemos associar ao movimento (órbita) de um corpo nas
proximidades da Terra, ou de outro corpo celeste, uma aceleração centrípeta que corresponderá ao
conceito clássico de \emph{aceleração devida à gravidade}.


\section{Definição operational de força peso do corpo}


O que os seres humanos e a restante matéria experimentam como ``peso" não é a força da gravidade. A
sensação de ``peso" é devida à força normal que os suportes exercem nos corpos, constrangindo-os de
forma a contrariar a força (``peso") que eles exercem nos suportes. Uma evidência deste facto
ocorre quando uma pessoa em cima de uma balança-dinamómetro verifica que o valor indicado por esta
(a intensidade do ``peso") varia sempre que flecte as pernas sem perder o contacto com a superfície
da balança. Outra manifestação ocorre quando um elevador a descer, trava para parar: uma pessoa
sente um acréscimo de pressão nas pernas e nos pés. Estas variações não podem ser atribuídas à
força da gravidade, porque a distância entre os centros de massa da Terra e da pessoa praticamente
não se alterou, assim como as respectivas massas.

\subsection{Força peso de um corpo em repouso}

\begin{figure}[hbt]
\begin{center}
\epsfbox{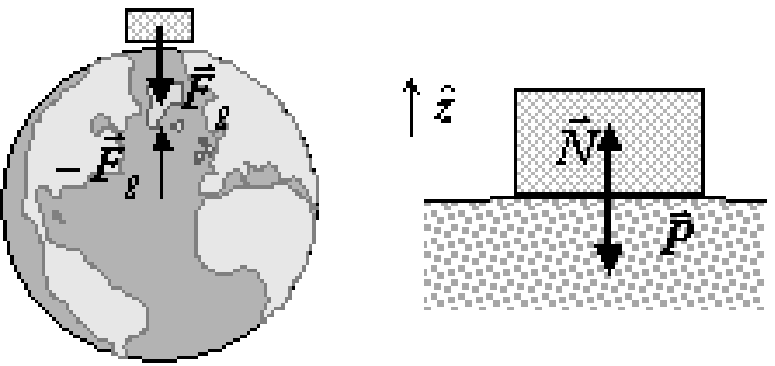}
\end{center}
\caption{\label{Fig2} a) \(\vec{F}_{g}\) e \(-\vec{F}_{g}\): par acção-reacção devido à interacção
gravitacional entre o corpo e a Terra. b) \(\vec{P}\) e \(\vec{N}\): par acção-reacção em resultado
da interacção entre as superfícies do corpo e da Terra.}
\end{figure}

Consideremos um corpo em repouso à superfície da Terra, Fig. \ref{Fig2}a. O corpo está sujeito à
força \(\vec{F}_{g}\) devido ao ``puxão" gravitacional da Terra. A força de reacção a esta força é
\(-\vec{F}_{g}\) e corresponde à acção gravitacional
exercida pelo corpo na Terra. O par de forças \(\vec{F}_{g}\) e \(-\vec{F}_{g}\) 
 constitui um par acção-reacção. A tendência do corpo em acelerar no sentido do centro da Terra devido à força \(\vec{F}_{g}\) dá
origem à força \(\vec{P}\), Fig. \ref{Fig2}b, força que a superfície do corpo exerce (acção do
corpo) na superfície da Terra.
Se o corpo exerce na superfície da Terra a força \(\vec{P}\), a superfície sólida da Terra reage
exercendo uma força \(\vec{N}\) no corpo que contrabalança a força
\(\vec{P}\), Fig. \ref{Fig2}b. A força \(\vec{N}\) 
designa-se força normal e é a reacção a \(\vec{P}\): \(\vec{P}\) e \(\vec{N}\) constituem um par
acção-reacção, \(\vec{N}=-\vec{P}\). A acção que o corpo exerce na superfície da Terra (ou suporte)
corresponde à definição operacional da força \textbf{``peso do corpo"}.

Em conclusão, no corpo actuam duas forças, a força \(\vec{F}_{g}\) apontando para o centro da Terra
e a força \(\vec{N}\) no sentido oposto. Uma vez que o corpo está em repouso relativamente à Terra
é nula a resultante das forças que actuam no corpo. Em consequência, as forças \(\vec{F}_{g}\) e
\(\vec{N}\) que actuam no corpo - em pontos de aplicação diferentes - têm a mesma intensidade e
orientações opostas. De forma similar, as duas força que actuam na Terra, \(-\vec{F}_{g}\) e
\(\vec{P}\), têm também a mesma magnitude, \(|\vec{P}|=|-\vec{F}_{g}|\), sentidos opostos e pontos
de aplicação diversos.

Consideremos agora o corpo pendurado num dinamómetro ou colocado no prato de uma
balança-dinamómetro. Quando o corpo é colocado na balança, a mola da balança é comprimida (no caso
do dinamómetro seria distendida) e a sua deformação é comunicada ao ponteiro da escala da balança.
O corpo exerce uma acção no prato da balança e através deste na mola. O ponteiro indica a magnitude
da força ``peso do corpo" \(\vec{P}\), força exercida \textbf{na} superfície do prato da balança
\textbf{pela} superfície do corpo.

Se o corpo e a balança estiverem em repouso em relação à Terra então, e como anteriormente, a
intensidade da força ``peso do corpo", que actua na balança, iguala a magnitude da força da
gravidade que actua no corpo, \(|\vec{P}|=|\vec{F}_{g}|\); a igualdade verifica-se também no caso
do corpo e da balança estarem em movimento uniforme e rectilíneo relativamente a um referencial
ligado à Terra.

Será que a relação \(|\vec{P}|=|\vec{F}_{g}|\) permanece válida se o corpo e o seu suporte
estiverem em movimento acelerado em relação à Terra? Qual é a indicação da balança nesta situação?

\subsection{Força peso de um corpo num referencial acelerado}

Imaginemos o corpo colocado no prato de uma balança fixa no pavimento de um elevador. Quando o
elevador acelera com uma aceleração \(\vec{a}\), enquanto estiver em contacto com a balança o corpo
move-se em conjunto com a balança e o elevador. No corpo actuam duas forças: a força da gravidade
\(\vec{F}_{g}\) e a reacção \(\vec{N}\) da balança à força ``peso do corpo" \(\vec{P}\). De acordo
com a segunda lei de Newton da Dinâmica e assumindo que a massa \(m\) do corpo não varia, a
resultante das forças que actuam no corpo é igual ao produto da massa \(m\) com a aceleração
\(\vec{a}\) do corpo, isto é,
\begin{equation}\label{000}
 m\vec{a}=\vec{F}_{g}+\vec{N}.
\end{equation}
Uma vez que a força ``peso do corpo" \(\vec{P}\) e a reacção \(\vec{N}\) da superfície da balança
formam um par acção-reacção, \(\vec{P}=-\vec{N}\), podemos rescrever a equação \ref{000} como
\begin{equation}\label{00}
  \vec{P}=m\vec{g}-m\vec{a}.
\end{equation}
Ter presente que o valor indicado na balança corresponde à intensidade da força \(\vec{P}\) que o
corpo exerce na balança.

Correntemente, define-se vertical de um lugar como a direcção da aceleração devida à gravidade.
Outras definições baseiam-se na direcção do fio-de-prumo, que à superfície da Terra e em repouso
coincide com a direcção da força da gravidade. A noção de vertical é mais geral. Por exemplo, um
ser humano ou outro ser vivo sente-se equilibrado na direcção da força que exerce no suporte e a
orientação ``para baixo" correspondem ao sentido dessa força. A vertical está sempre segundo a
linha de acção da força ``peso do corpo" e a orientação ``para baixo" corresponde ao sentido dessa
força. Da relação \ref{00} pode concluir-se que tanto a direcção da vertical de um corpo como as
orientações ``para baixo/cima" dependem da aceleração do corpo e da aceleração devida à gravidade.
Uma constatação quotidiana ocorre quando um autocarro arranca ou quando trava: nesta situação a
``nossa" vertical é oblíqua e para não nos desequilibrarmos inclinamo-nos na direcção da ``nova"
vertical.


\section{Força peso do corpo num campo gravitacional fraco}

A acção que um corpo exerce nos seus suportes, a força ``peso do corpo", não depende da existência
de um campo gravitacional na região do espaço onde este se encontra. Considere-se um veículo
espacial numa região do Universo onde o efeito gravitacional é nulo ou pouco significativo, i.e.,
\(\vec{g}\simeq\vec{0}\). Se uma nave em movimento rectilíneo e uniforme nesta região accionar os
seus motores entrará em movimento acelerado e os objectos no seu interior serão ``projectados" no
sentido oposto à aceleração \(\vec{a}\) da nave, tal como acontece quando um carro acelera,
acabando estes por exercer forças nos seus suportes ou nas paredes da nave. Isto é, quando a nave
espacial acciona os motores passa a ser actuada pela força de propulsão que é transferida a cada
objecto no seu interior. Estes, após entrarem em contacto com as paredes da nave ou com outros
corpos solidários com a nave, deslocam-se com a aceleração \(\vec{a}\) da nave. Nestas condições a
equação \ref{00} toma a forma
\begin{equation}\label{0000}
    \vec{P}=-m\vec{a},
\end{equation}
i.e., a acção que o corpo exerce no seu suporte tem o sentido oposto à aceleração da nave e depende
apenas da intensidade dessa aceleração e da massa do corpo.

Pode, portanto, concluir-se que a força que os corpos exercem nos seus suportes ou a ``sua
ausência" não têm a ver necessariamente com o facto do corpo estar ou não sob a influência de um
campo gravitacional. Mesmo na presença de um campo gravitacional significativo, esta força depende
essencialmente das características de movimento do corpo e do seu suporte. De facto, as propostas
para ``criar artificialmente" o efeito da gravidade no interior de naves interplanetárias empregam
a rotação da nave em torno do seu centro de massa, sendo a aceleração devida à ``gravidade
artificial" oposta à aceleração centrípeta da nave.

\section{Imponderabilidade e microgravidade}



Nas situações ideais de queda livre todas as partes de um avião ou de nave espacial acelerariam
uniformemente e o ambiente no seu interior seria de ``gravidade-zero", porque os corpos no seu
interior não sentiriam os efeitos da gravidade. Em situações reais o efeito da gravidade nos
corpos, embora substancialmente reduzido, faz-se sentir, e diz-se que os corpos estão em ambiente
de ``microgravidade" porque a ``ausência de peso" não é total. Contudo, o emprego do termo
``microgravidade" sem especificar o seu significado real pode aumentar os equívocos associados à
identificação da força ``peso" com a força da gravidade.

A total ``ausência de peso" numa nave espacial só seria possível se todos os pontos da nave
estivessem em movimento rectilíneo e uniforme, fora da acção de qualquer campo gravitacional. Neste
caso, os corpos dentro da nave não experimentariam qualquer aceleração uns relativamente aos
outros, e a situação seria de gravidade-zero, desprezando é claro as interacções gravitacionais
mútuas e com as paredes da nave.
As situações de imponderabilidade (``ausência de peso") correntes são frequentemente designadas, de
forma imprópria, como ``gravidade-zero" ou ``microgravidade". É importante ter presente que no
interior de uma nave nas proximidades da Terra a intensidade da aceleração devida à gravidade
terrestre não é nula (``gravidade-zero") ou diminuta (``microgravidade"): a 400 km de altura, por
exemplo, a aceleração devida à gravidade terrestre é 8,4 m s\(^{-2}\), e mesmo à distância da
órbita da Lua é 2,6\(\times10^{-3}\) m s\(^{-2}\), embora nessas regiões a aceleração devida à
gravidade solar seja cerca de duas vezes superior, i.e., aproximadamente 5,8\(\times10^{-3}\) m
s\(^{-2}\). A verdadeira microgravidade terrestre, \(g\approx 10^{-6}\) m s\(^{-2}\), só poderia
ser experimentada em regiões à distância de 17 vezes a separação entre a Terra e a Lua.

Para os cientistas, o termo microgravidade caracteriza o facto de que a nave em queda livre e os
corpos no seu interior não estarem todos sujeitos exactamente à mesma aceleração, em resultado das
diferentes interacções gravitacionais entre os corpos interiores e exteriores à nave, da acção de
outras forças, e traduz-se na aceleração residual que os corpos no interior de um veículo em queda
livre experimentam relativamente, por exemplo, ao centro de massa deste. Em consequência, nas
proximidades da Terra e em naves não tripuladas em queda livre obtém-se com facilidade \(1 \mu g\),
enquanto que em missões tripuladas dificilmente se consegue menos de \(100 \mu g\). As causas mais
relevantes são: i) a variação da aceleração devida à gravidade em resultado da morfologia da Terra;
ii) variações devidas ao efeito gravitacional dos outros corpos celestes, em particular do Sol e da
Lua, dependentes das posições relativamente à Terra
iii) a variação da aceleração da gravidade terrestre com a altitude que decresce aproximadamente 1
parte por milhão por cada 3 m de aumento (numa nave em órbita a força centrípeta e, portanto, a
aceleração devida à gravidade é superior na parte da nave mais afastada da Terra do que na parte
mais próxima);
iv) a atmosfera, embora podendo ser muito rarefeita, por exemplo, a 400 km de altura, desacelera
gradualmente a nave.

\section{Conclusão}

A identificação da força ``peso" com a força da gravidade carece de lógica do ponto de vista do
conhecimento actual e é, muitas vezes, fonte de confusão.
Na definição operacional discutida neste texto, a força ``peso do corpo" corresponde à acção que
este exerce no seu suporte, e depende do estado de movimento de ambos. Tendo presente que o
conceito de ``peso" não é fundamental em Física, acredita-se que a aprendizagem desta ciência sairá
beneficiada se a expressão ``peso do corpo" não for usada. Uma vantagem óbvia é eliminar a confusão
frequente na disciplina entre o conceito de massa do corpo e de força ``peso do corpo".
Espera-se que este texto ajude a motivar a comunidade científica, os professores e os autores dos
manuais de Física, a adoptarem a definição operacional da força ``peso", embora o abandono do
enunciado força ``peso do corpo" seja desejável.



\section*{Agradecimentos}

O autor agradece aos Professores Robertus Potting, Paulo Sá, José Rodrigues, e Paulo Silva os
comentários e a revisão do manuscrito.
